\documentclass[12pt]{iopart}
\usepackage{graphicx}
\usepackage[utf8]{inputenc}
\usepackage{amssymb}
\usepackage{setstack}

\begin{document}
\title{Negative spin polarization of Mn$_2$VGa probed by tunnel magnetoresistance} 
\author{Christoph Klewe, Markus Meinert, Jan Schmalhorst, G\"unter Reiss}
\address{Department of Physics, Bielefeld University, 33501 Bielefeld, Germany}
\ead{cklewe@physik.uni-bielefeld.de}
\date{\today}

\begin{abstract}
The ferrimagnetic Heusler compound Mn$_2$VGa is predicted to have a pseudogap in the majority spin channel, which should lead to a negative tunnel magnetoresistance. We synthesized epitaxial Mn$_2$VGa thin films on MgO$(001)$ substrates by dc and rf magnetron co-sputtering, resulting in nearly stoichiometric films. XRD analysis revealed a mostly B2-ordered structure for the films deposited at substrate temperatures of $350^\circ$C, $450^\circ$C, and $550^\circ$C. Magnetic tunnel junctions with MgO barrier and CoFe counter-electrodes were fabricated. After post-annealing at up to T$_\mathrm{a}=425^\circ$C negative TMR was obtained around zero bias, providing evidence for the inverted spin-polarization. Band structures of both electrodes were computed within the coherent potential approximation and used to calculate the TMR($V$) characteristics, which are in good agreement with our experimental findings.

\end{abstract}

\submitto{\JPD}

\maketitle

\section{Introduction}
Since their discovery by de Groot et al.\cite{deGroot83}, half-metallic magnets have become a subject of growing interest. This new class of materials shows metallic transport properties for spins of one orientation, while being insulating for electrons with opposite spin orientation, because a band gap at the Fermi level $\varepsilon_{\mathrm{F}}$ is present.  Usually, the density of states (DOS) of the majority-spins has a finite value at $\varepsilon_{\mathrm{F}}$. This results in a spin polarization of the conduction electrons of $100\,\%$.
 
Promising candidates to show half-metallic character and high spin polarization are the full Heusler compounds.\cite{Galanakis02}
This class of materials is characterized by the formula $X_{2}YZ$, where $X$ and $Y$ represent transition metals and $Z$ is a main group element. Full Heusler compounds crystallize in the L$2_{1}$ structure, a face-centered cubic structure with a four atom basis. The basis vectors are $A(0,0,0)$, $B(\frac{1}{4},\frac{1}{4},\frac{1}{4})$, $C(\frac{2}{4},\frac{2}{4},\frac{2}{4})$, and $D(\frac{3}{4},\frac{3}{4},\frac{3}{4})$.
The $X$ elements occupy the $A$ and $C$ sites, while the $Y$ and $Z$ elements arrange on the $B$ and $D$ sites, respectively.
Chemical disorder can affect the electronic properties significantly.
The B$2$ structure with an intermixing of $Y$ and $Z$ elements on the $B$ and $D$ sites occurs frequently. 

Half-metallic ferrimagnetic compounds show advantages as compared to the more common half-metallic ferromagnets. Due to their internal spin compensation they have a lower magnetization and thus do not generate high stray fields, while still having high Curie-temperatures.\cite{Pickett01} 

Recently, the class of Mn$_2$V$Z$ Heusler compounds has been the subject of several theoretical studies.\cite{Sasioglu05, Oezdogan06} \textit{Ab initio} calculations performed for compounds with $Z$ = Al, Ga, In, Si, Ge, and Sn by \"Ozdogan \textit{et al.}\cite{Oezdogan06} predicted a nearly half-metallic electronic structure with a pseudo band gap for the majority carriers and a ferrimagnetic ground state for most of these compounds. Furthermore, they follow the Slater-Pauling rule for the magnetic moment in half-metallic compounds.\cite{Galanakis02} The best known compound from this class is Mn$_2$VAl, where good agreement between experiments\cite{Yoshida81, Kawakami81, Nakamichi83, Itoh83, Jiang01, Kubota09, Meinert11JPD} and theory\cite{Sasioglu05, Oezdogan06, Ishida84, Weht99, Galanakis07, Chioncel09} is achieved.

In contrast to the well-known Co$_2$-based Heusler compounds with a minority gap, the majority gap results in a negative spin polarization, and thus negative tunnel magnetoresistance (TMR) in magnetic tunnel junctions (MTJs) around zero bias is expected. However, the presence of such an inverted spin polarization remains to be observed experimentally.

In this study, we investigate thin films of the full Heusler compound Mn$_2$VGa. We demonstrate that epitaxial growth can be achieved by magnetron sputtering and discuss the structural, magnetic, and electronic transport properties of the films. In particular, magnetic tunnel junctions are investigated and the inverse spin polarization is discussed based on an \textit{ab initio} model of the electronic transport.

\subsection{Experimental details}
Mn$_2$VGa thin films were produced by ultra high vacuum dc and rf magnetron co-sputtering. The base pressure in the sputter chamber was better than $10^{-8}\,\mathrm{mbar}$. The target-to-substrate distance was 21\,cm, and the substrate was rotated with 10 rounds per minute to obtain homogenous films.

Due to the low sticking coefficient of Ga, we obtained nearly stoichiometric Mn$_2$VGa thin films by using a Mn$_{50}$Ga$_{50}$ alloy target and an elemental V target.
X-ray reflectivity (XRR) and x-ray fluorescence (XRF) were used to calibrate the sputter parameters and to analyse the stoichiometry of the fabricated samples. The measurements revealed a small overweight of Mn with a Mn\,:\,Ga ratio of 2.2\,:\,1. By adapting the V ratio, we obtained the stoichiometry Mn$_{2.047}$V$_{1.023}$Ga$_{0.93}$. In this composition, the V content was adjusted to match the Mn content in the correct 2\,:\,1 proportion.

The samples were deposited on MgO(001) substrates with a lattice parameter of 4.21\,\AA{}. The lattice mismatch between the substrate and bulk Mn$_2$VGa ($a = 5.905$\,\AA{})\cite{Kumar08} is less than 1\%. In order to determine the best deposition parameters, $20\,\mathrm{nm}$ films were fabricated at substrate temperatures $T_\mathrm{S}$ of room temperature (RT), $350^\circ$C, $450^\circ$C, and $550^\circ$C. 
The layers were capped with $2\,\mathrm{nm}$ MgO to prevent oxidation. These samples are referred to as half-stacks.

Crystallographic properties were investigated by means of x-ray diffraction (XRD) in a Philips X'Pert Pro diffractometer with a Cu anode and a Bragg-Brentano arrangement. Quantification of chemical disorder was performed with off-specular measurements on an open Euler cradle and point focus collimator optics.

Magneto-optic Kerr effect (MOKE) and vibrating sample magnetometry (VSM) were used to investigate the magnetic properties. VSM measurements were performed at 50 and 300\,K.

X-ray magnetic circular dichroism (XMCD) was measured at room temperature at beamlines 4.0.2 and 6.3.1 of the Advanced Light Source (ALS), Berkeley. 

To examine the transport properties we fabricated full magnetic tunnel junctions with a $20\,\mathrm{nm}$ Mn$_{2.047}$V$_{1.023}$Ga$_{0.93}$ bottom electrode deposited at $450^\circ$C. A $2.5\,\mathrm{nm}$ thick MgO tunnel barrier was fabricated by electron beam evaporation. 

The samples were patterned to junctions of $7.5\times 7.5 \, \mu \mathrm{m}^2$ by UV lithography.
Room temperature measurements of the tunnel magnetoresistance were performed in the as prepared state and after succesive post annealing treatments up to $425^\circ$C with a two-point probe technique in a magnetic field of up to $3\,\mathrm{kOe}$.

\subsection{Computational approach}
To enable a detailed understanding of our experimental results, we model the electronic transport properties on the basis of \textit{ab initio} density functional theory calculations.

The calculations of ordered and chemically disordered compounds were performed with the \textit{Munich} SPR-KKR package, a spin-polarized relativistic Korringa-Kohn-Rostoker code.\cite{SPRKKR} The ground state self-consistent calculations were performed on $28\times28\times28$ \textbf{k} point meshes in the irreducible wedge of the Brillouin zone. The exchange-correlation functional was approximated by the Perdew-Burke-Ernzerhof implementation of the generalized gradient approximation.\cite{PBE} The angular momentum expansion was taken up to the maximum available value $l_\mathrm{max} = 3$. To improve the charge convergence with respect to the angular momentum cutoff, the Fermi energy was determined using Lloyd's formula.\cite{Lloyd72,Zeller08}  A scalar relativistic representation of the valence states was used in all cases, thus neglecting the spin-orbit coupling. The experimental bulk lattice parameters (5.905\,\AA{} for Mn$_2$VGa\cite{Kumar08} and 2.85\,\AA{} for Co$_{70}$Fe$_{30}$) were adopted for the calculations. To account for chemical disorder, the coherent potential approximation (CPA) was used.

For comparison, we have performed calculations with the highly accurate full-potential linearized augmented plane-wave (FLAPW) method, implemented in the Elk code.\cite{elk} The muffin-tin radii were set to 2.3\,bohr, and angular momentum and plane-wave cutoffs of $l_\mathrm{max} = 9$ and $k_\mathrm{max}=3.7\,\mathrm{bohr}^{-1}$ were used.

\section{Results}

\subsection{Crystal structure and magnetic properties}
\begin{figure}[t]
	\centering
		\includegraphics{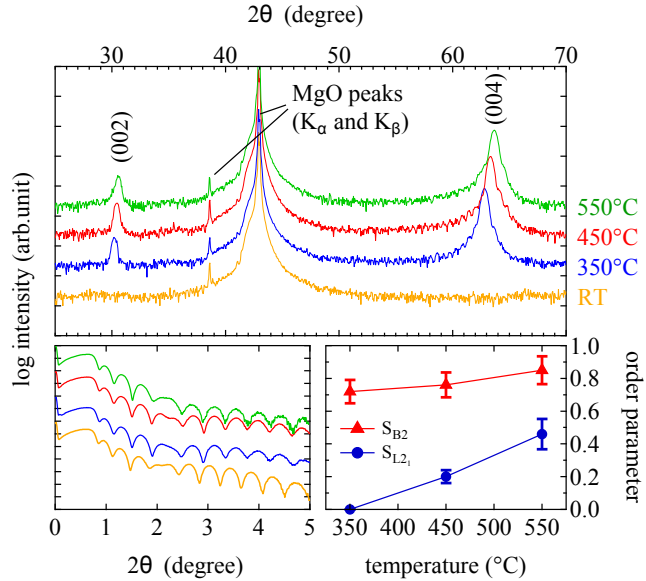}
	\caption{(Color online) XRD- (top) and XRR- (bottom left) spectra of Mn$_2$VGa for different deposition temperatures. Bottom rigth: L2$_1$ and B2 order parameters for different deposition temperatures.}	
	\label{fig:XRD_XRR}
\end{figure}

The X-ray diffraction and reflectivity measurements of the Mn$_2$VGa films are presented in Figure \ref{fig:XRD_XRR}.
The film deposited at room temperature shows a smooth surface as visible in the pronounced reflectivity oscillations but is not crystallized.
For temperatures above $350^\circ$C the films exhibit crystallization and highly (001) oriented grains, as can be derived from very sharp $(004)$ rocking-curves with FWHM values  of $0.1^\circ$(not shown). Laue-oscillations at the $(004)$ peaks are visible in the XRD patterns, displaying the good crystallization. The reflectivities indicate that the $550^\circ$C sample possesses a higher roughness of about $0.7\,\mathrm{nm}$, while the films deposited at $350^\circ$C and $450^\circ$C are very smooth with roughnesses of only $0.1\,\mathrm{nm}$. 
Lattice parameters perpendicular to the surface decrease with increasing deposition temperature. The $350^\circ$C film has the highest value of about $5.913\, \mathrm{\AA}$, which is a little larger than the bulk value. The lattice parameters $5.891\, \mathrm{\AA}$ and $5.842\, \mathrm{\AA}$ in the $450^\circ$C and the $550^\circ$C samples, respectively, are significantly lower. 
This trend in lattice distortion is due to an increasing lattice matching between the MgO(001) substrate and the films with increasing temperature. The lattice thus is expanded in the film plane and compressed in the direction perpendicular to the plane.

The deposition temperature 450$^\circ$C was chosen for the MTJs since these films had sufficiently small roughness, allowing to grow tunnel barriers free of pinholes.

The order parameters were determined with the method presented by Takamura \textit{et al.}\cite{Takamura08} We found a mostly B2 ordered structure in the films with only a small degree of L2$_1$ ordering.
With increasing deposition temperature the order parameters increased. 
Especially, an increase in L2$_1$ ordering at higher deposition temperatures occured.
The order parameters, which have been obtained including anomalous scattering contributions are displayed in Figure \ref{fig:XRD_XRR}. Due to the similar atomic form factors of Mn and V, the parameter $S_\mathrm{B2}$ measures essentially the number of Ga atoms on the Mn sites. With $S_\mathrm{B2} = 0.8$ we find that about 10\% of the Mn sites are occupied with Ga atoms. Assuming that the A, B, and C sites are ideally occupied and the excess Mn and V of our samples reside on Ga sites, we have $S_\mathrm{B2} \leq 0.91$. Hence, some additional disorder with Ga on A, C sites is present in the samples.
At temperatures higher than $550^\circ$C, substantial evaporation of Mn from the hot substrate is observed and it becomes impossible to control the stoichiometry.

\begin{figure}[t]
	\centering
		\includegraphics{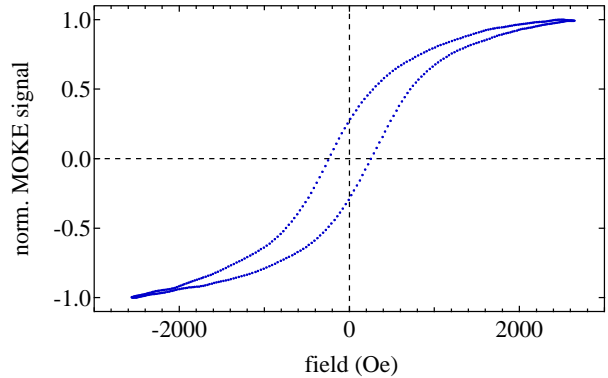}
	\caption{MOKE measurement on a film deposited at $450^\circ$C.}
	\label{fig:MOKE}
\end{figure}

In-plane MOKE measurements showed a ferromagnetic behaviour with wide sigmoidal hysteresis loops, shown in Figure \ref{fig:MOKE}.
The coercivities were in the range of $125\,\mathrm{Oe}$ to $195\,\mathrm{Oe}$ with squarenesses between 0.1 and 0.18.

The extrapolated low-temperature magnetic moment of the films deposited at 450$^\circ$C was $(1.56 \pm 0.03)\,\mu_\mathrm{B}$\,/\,f.u.. At room temperature, this value is reduced to $(1.31 \pm 0.03)\,\mu_\mathrm{B}$\,/\,f.u., which is consistent with a Curie temperature of more than 700\,K, as was determined for bulk samples.\cite{Kumar08}

\subsection{X-ray absorption spectroscopy and circular dichroism}
\begin{figure}[t]
	\centering
		\includegraphics{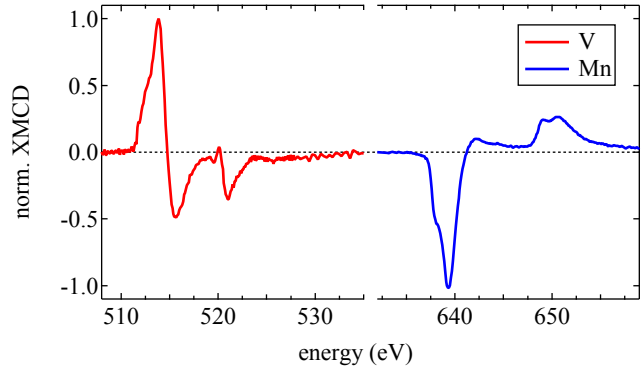}
	\caption{XMCD measurements at the L$_{2,3}$ edges of V and Mn for a sample deposited at $550^\circ$C. Taken at beamline 4.0.2 of the ALS in luminescence detection.}
	\label{fig:XMCD}
\end{figure}

In Fig. \ref{fig:XMCD} we show exemplarily the L$_{2,3}$ XMCD spectra of a sample deposited at 550$^\circ$C. The spectra of the other samples are very similar. The antiparallel alignment of the Mn and V magnetic moments is immediately clear. Unfortunately, we can not quantify the magnetic moment of V. On the one hand, this is because we can not determine the experimental background function, due to the K absorption edge of O at 543\,eV. On the other hand, jj mixing prevents us from applying the sum rules in the usual way. To compensate for the latter in the case of Mn, we multiply the apparent spin moment with 1.5.\cite{Duerr97} For the 550$^\circ$C sample we find $m_\mathrm{Mn}^s = (1.26 \pm 0.25) \, \mu_\mathrm{B}$ per atom at room temperature, where the error arises mainly from uncertanties in the background subtraction. This value extrapolates to $m_\mathrm{Mn}^s = (1.50 \pm 0.3) \, \mu_\mathrm{B}$ per atom at low temperature, which is consistent with the calculated magnetic moment (see next section).

Neutron diffraction performed by Itoh \textit{et al.} on the isostructural and isoelectronic compound Mn$_2$VAl revealed a ferrimagnetic coupling of Mn and V with total moments of $m_\mathrm{Mn} = (1.50 \pm 0.3) \, \mu_\mathrm{B}$ and $m_\mathrm{V} = -0.9 \, \mu_\mathrm{B}$ at room temperature.\cite{Itoh83} These data are consistent with a total magnetic moment of 2\,$\mu_\mathrm{B}$\,/\,f.u. for the ideally ordered compound, and they are consistent with our measurements on Mn$_2$VGa, which should have similar magnetic moments. We note that the XMCD spectra shown here are very similar to those of Mn$_2$VAl.\cite{Meinert11JPD}

\subsection{Magnetic ground state and structural model}

In our SPR-KKR calculation with the ideally ordered L2$_1$ structure, Mn$_2$VGa has a total moment of 1.94\,$\mu_\mathrm{B}$/f.u., with 1.37\,$\mu_\mathrm{B}$ on Mn and -0.76\,$\mu_\mathrm{B}$ on V. In the FLAPW calculation, the total magnetic moment is 1.98\,$\mu_\mathrm{B}$/f.u., with 1.47\,$\mu_\mathrm{B}$ on Mn and -0.91\,$\mu_\mathrm{B}$ on V. Due to the limited accuracy with respect to the angular momentum cutoff in the SPR-KKR calculation, we consider the FLAPW result to be more accurate. Furthermore, these values match the experimental values of the magnetic moments of Mn$_2$VAl,\cite{Itoh83} for which we find very similar magnetic moments as for Mn$_2$VGa. Similar results were also obtained for the Mn$_2$Ti\textit{Z} compounds.\cite{Meinert2011JPCM}

However, chemical disorder is more accurately treated in the KKR method, as supercells have to be considered in the FLAPW scheme. Thus, we treat the chemical disorder within the KKR method, noting that the magnetic moments are eventually somewhat underestimated. In both calculations, the band structure is not fully half-metallic, but a pseudogap for the majority spin is present at the Fermi energy.

The B2 ordered case exhibits a similar pseudogap as the L2$_1$ ordered compound. Here, the total moment is 1.93\,$\mu_\mathrm{B}$/f.u., with 1.42\,$\mu_\mathrm{B}$ on Mn and -0.87\,$\mu_\mathrm{B}$ on V. The total magnetic moment does not depend on the degree of V-Ga intermixing, i.e., it is independent from $S_\mathrm{L2_1}$. 

However, we find a significant influence of disorder involving Mn. An L2$_1$ structure with 5\,\% Mn and 3\,\% V antisites on the Ga sites mimics the experimental stoichiometry. The calculated total magnetic moment of this configuration is 1.71\,$\mu_\mathrm{B}$\,/\,f.u., where the individual magnetic moments correspond to those of the ideal L2$_1$ structure. The Mn and V antisites have -2.37\,$\mu_\mathrm{B}$ and -0.88\,$\mu_\mathrm{B}$, respectively. The calculated total moment agrees with the Slater-Pauling value of 1.74\,$\mu_\mathrm{B}$\,/\,f.u. for this stoichiometry. Nevertheless, both values are larger than the experimental magnetic moment, which hints to additional Mn disorder being present in the samples, as was already suggested by the XRD measurements.

With an additional 4\% intermixing of Mn and Ga, considering the Mn antisite moments as calculated above, we get $S_\mathrm{B2} = 0.82$ and $m_\mathrm{tot}^\mathrm{s} = 1.56$\,$\mu_\mathrm{B}$/f.u., which agree with the experimental data. The total Mn occupation of Ga sites is 9\% within this model. A small amount of  additional Mn-V disorder would not influence the total magnetic moment: the V antisites would couple parallel to the Mn atoms on the (A,C) sites and partially compensate the large negative moment of the Mn antisites. Assuming the moments to be independent from the degree of disorder, Mn-Ga swaps contribute about $-3.8\,\mu_\mathrm{B}$ per atom, while Mn-V swaps give only about $-2\,\mu_\mathrm{B}$ per atom. However, with Mn-V swaps the Mn (A,C) moments increase slightly in addition, so that in total the moment remains nearly constant. This type of disorder can not be detected by XRD, so Mn-V disorder remains unknown.

These considerations establish the following internally consistent structural model for our films. The ordering is essentially of the B2 type, with some preferential ordering of the V and Ga atoms towards the L2$_1$ structure. In addition, the (V,Ga) sites have a 4.5\% occupation of Mn atoms due to excess of Mn atoms and chemical disorder involving Ga. The Mn antisites couple antiparallel to the Mn atoms on the (A,C) sites and reduce the total magnetic moment.

\subsection{Electronic structure and transport properties}
\begin{figure}[t]
	\centering
		\includegraphics{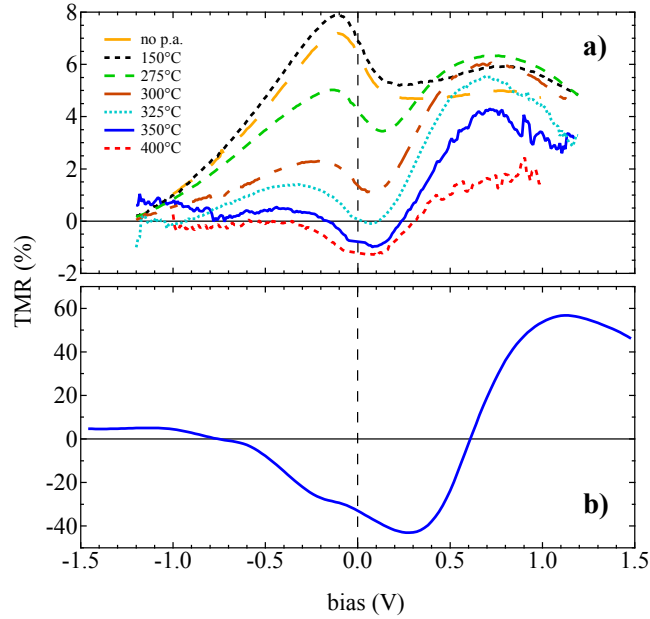}
	\caption{(Color online) a): Experimental TMR($V$) characteristics at RT after succsessive post-annealing.
	b): Calculated TMR($V$) curve for \textit{s} electron transport in the B2 ordered system.}	
	\label{fig:TMRvsU}
\end{figure}

To investigate the transport properties of the magnetic tunnel junctions, major loops and $I(V)$ curves were recorded.
The TMR($V$) characteristics were determined from the $I(V)$ curves for parallel and antiparallel magnetization.
We carried out measurements on the samples without and after successive post-annealing treatments, starting at 100$^\circ$C and up to 425$^\circ$C in steps of 25$^\circ$C each for 30\,min. The results are displayed in Figure \ref{fig:TMRvsU}a) for selected temperatures T$_\mathrm{a}$.

\begin{figure}[t]
	\centering
		\includegraphics{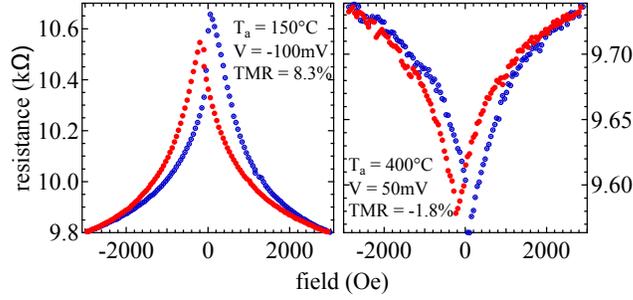}
	\caption{Left: Positive tunnel magnetoresistance of $8.3\,\%$ at $-100\,\mathrm{mV}$ after annealing at 150$^\circ$C.
	Right: Negative tunnel magnetoresistance of $-1.8\,\%$ at 50\,mV after annealing at 400$^\circ$C.}	
	\label{fig:TMR}
\end{figure}

A distinct dependency of TMR($V$) on the annealing temperature is visible. The highest TMR was obtained at a bias voltage of -100\,mV after annealing at 150$^\circ$C, where positive bias means tunneling of electrons into the Heusler compound. Major loops (Figure \ref{fig:TMR}) yield a TMR ratio of 8.3$\,\%$ for these parameters.  However, the field of $\pm 3000\,\mathrm{Oe}$ is not high enough to saturate the  Mn$_2$VGa completely, see Figure \ref{fig:MOKE}. Futhermore, the low squareness of the Mn$_2$VGa hysteresis reduces the TMR by at least a factor of 2.

The 150$^\circ$C curve also shows a shoulder at high positive bias voltages between $+500$ and $+1000\,\mathrm{mV}$.
With increasing temperature the shape of this shoulder develops to a peak while the maximum around $-100\,\mathrm{mV}$ is reduced significantly and a minimum around +100\,mV emerges.
As of T$_\mathrm{a}=325^\circ$C, the tunnel magnetoresistance drops below zero for low positive bias voltages.
For higher post-annealing temperatures, the TMR around low bias is significantly negative. This is an indication for the inverted spin polarization and the pseudogap in the majority channel.
The highest negative TMR of \mbox{$-1.8\, \%$} was reached for T$_\mathrm{a}=400^\circ$C at V$=+50\,$mV (Figure \ref{fig:TMR}).

\begin{figure}[t]
	\centering
		\includegraphics{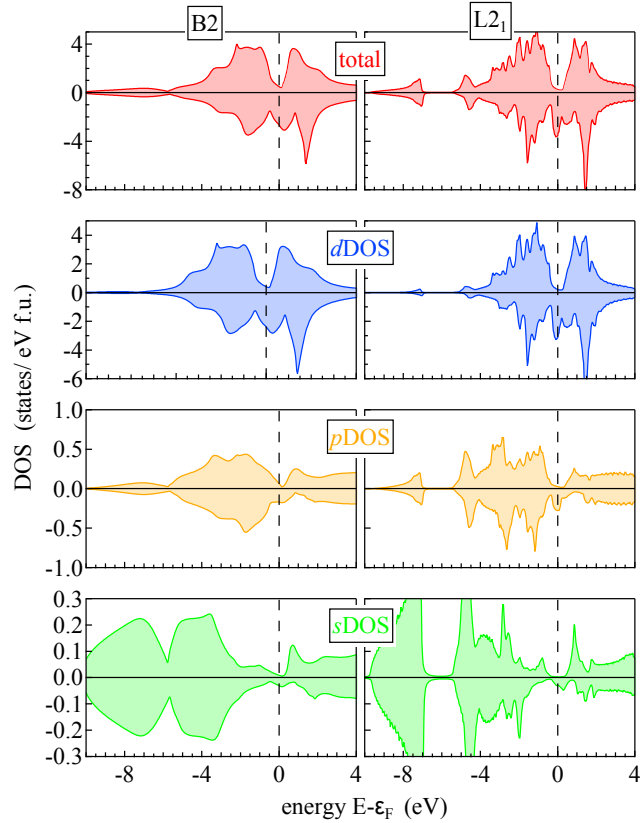}
	\caption{(Color online) Angular momentum projected Mn$_2$VGa density of states calculated for a B2 (left column) and an L2$_1$ ordered (right column) system. Spin majority is shown as positive, spin minority is shown as negative.}	
	\label{fig:DOS}
\end{figure}

To provide an explanation for the bias voltage dependence of the TMR, we consider the spin- and angular momentum resolved densities of states of the B2 ordered Mn$_2$VGa compound, see Figure \ref{fig:DOS}. For comparison, DOS of the L2$_1$ structure are shown along with the B2 graphs. We neglect bands with \textit{f} character.
Both for the B$2$ and the L2$_1$ ordered structures the DOS consists almost completely of \textit{d} character states around the Fermi level (more than 80$\%$), while states with \textit{s} character contribute very little (less than 2$\%$). The spin polarization is negative in all cases. We find P$_\mathrm{total}\approx-69\,\%$ and P$_{s}\approx-55\,\%$ in the case of B2 order and about $-84\,\%$, both for the total DOS and the \textit{s} states exclusively in the L2$_1$ ordered case.

From the DOS we calculate the TMR($V$) characteristics based on Simmons' model:\cite{Simmons}
\begin{eqnarray}
I(V) \propto &\sum\limits_\sigma & \int _{-\infty}^{\infty} \rho_1^\sigma(E) \rho_2^\sigma(E+eV) \vert T(V) \vert ^2 \nonumber \\
&\times & f(E)[1-f(E+eV)]\mathrm{d}E,
\label{Tunnelstrom}
\end{eqnarray}
where $\rho_i^\sigma(E)$ denote the DOS of the two electrodes with spin $\sigma$ and $f(E)$ is the Fermi function. $\vert T(V) \vert ^2$ is the transmission probability and is proportional to $e^{-d\sqrt{\phi - V}}$ according to Stratton.\cite{Stratton1962} In this model, the tunnel magnetoresistance is independent of the barrier thickness $d$ and the potential height $\phi$.
The TMR can be calculated from the total currents for parallel and antiparallel orientation of the magnetizations.

The Co$_{70}$Fe$_{30}$ compound was modeled in the chemically disordered bcc (A2) phase. Its total moment is 2\,$\mu_\mathrm{B}$/f.u., with 1.76\,$\mu_\mathrm{B}$ on Co and 2.58\,$\mu_\mathrm{B}$ on Fe. As before, the density of states is dominated by \textit{d} states around $\varepsilon_\mathrm{F}$. The \textit{d} character and total DOS have negative spin polarization at $\varepsilon_\mathrm{F}$, while the \textit{s} states show a positive spin polarization.

Since our samples were primarily B2 ordered, we calculated the TMR($V$) curves for this case. A comparison between the theoretical and the experimental bias voltage dependent TMR is shown in Figure \ref{fig:TMRvsU}. The main contribution to the tunneling current comes from the light electrons of \textit{s} character.\cite{Hertz73, Stearns77} Thus, we consider only the \textit{s}DOS in Eq. \ref{Tunnelstrom}. With this model we find good qualitative agreement between experiment (for junctions with high post-annealing temperature) and theory: a negative TMR region around zero bias, a flat TMR($V$) at negative, and a peaked TMR(V) at positive bias, respectively. However, the voltage scale is narrower in the experiment. 

On the origin of the annealing temperature dependence of the TMR($V$) curves we can only speculate. Removal of interfacial disorder or reduction of oxidized transition metal atoms would provide explanations for such behaviour. Interfacial disorder could introduce states in the pseudogap, which would contribute to a positive TMR around zero bias. Oxidized transition metal atoms could allow coupling of \textit{d} electron states across the barrier (eventually as resonant defect states).\cite{Tsymbal07} Crystallization of the MgO barrier is usually expected to give rise to coherent tunneling and a strong enhancement of the TMR.\cite{Yuasa07, Ikeda08} In this case, only Bloch states with momentum parallel to the interface normal (here the $\Gamma-X$ direction) contribute to the tunneling current. However, this is inconsistent with our observation that the TMR($V$) characteristic is well explained with DOS obtained from integrating over the entire Brillouin zone.

The magnitude of the negative TMR of the $T_\mathrm{a} = 400^\circ$C sample is much smaller than the theoretical value of $-43\%$. On the one hand, this is due to the low squareness of the magnetic hysteresis. On the other hand, the theoretical value corresponds to a low-temperature experiment, while our measurements were taken at room temperature. Bulk disorder may in general induce states in the half-metallic gap.\cite{Picozzi03} We find that Mn antisites on Ga positions and vice versa contribute weakly to the pseudogap, but their polarization is negative for the \textit{s} states and in total. However, V antisites on Mn positions generate states in the pseudogap by hybridization, diminishing the negative polarization. Thus, the TMR($V$) dependence on the annealing temperature may also arise from removal of (interfacial) Mn-V disorder.

Inelastic tunneling contributions, such as magnon excitations or Kondo scattering, may also decrease the TMR. Furthermore, the interfacial spin polarization may be significantly different from the bulk polarization.\cite{Picozzi03}

It is well known that the tunneling spin polarization of the CoFe/MgO interface is positive.\cite{Parkin04} In conclusion, a negative spin polarization at the Mn$_2$VGa/MgO interface must be present in order to explain the observed negative TMR. The observed TMR($V$) characteristic is a definitive fingerprint of the majority spin pseudogap predicted by DFT calculations.

\section{Conclusions}
Epitaxial Mn$_2$VGa thin films were synthesized on MgO(001) substrates by UHV dc and rf magnetron co-sputtering. Crystallographic investigations revealed a mostly B2 ordered structure for all deposition temperatures with a small degree of L2$_1$ ordering. The order parameters are consistent with the magnetization, which is confirmed by \textit{ab initio} electronic structure calculations.

A series of successive post-annealing steps with TMR measurements on MTJs reveals a strong dependency of the TMR($V$) characteristics on the annealing temperature for this compound. For high annealing temperatures, we found negative TMR of up to $-1.8\,\%$ around low bias at room temperature. This can be assigned to the unusual band structure and reflects the inverse spin polarization with a pseudogap in the spin majority channel.

\section*{Acknowledgments}
The authors gratefully acknowledge financial support by the Deutsche Forschungsgemeinschaft (DFG, Contract RE1052/22-1)  and the Bundesministerium f\"ur Bildung und Forschung (BMBF). They thank for the opportunity to work at BL 6.3.1 and BL 4.0.2 of the Advanced Light Source, Berkeley, USA, which is supported by the Director, Office of Science, Office of Basic Energy Sciences, of the U.S. Department of Energy under Contract No. DE-AC02-05CH11231.
They acknowledge the work of Prof. Ebert and his group for developing and providing the \textit{Munich} SPR-KKR package. Furthermore they thank Tim B\"ohnert and Kornelius Nielsch from the University of Hamburg for VSM measurements.

\end{document}